# Noble metal free photocatalytic $H_2$ generation on black $TiO_2$: On the influence of crystal facets vs. crystal damage

Ning Liu[1], Hans-Georg Steinrück[2], Andres Osvet[3], Yuyun Yang[1,4], Patrik Schmuki[1,5*]

[1]*Department of Materials Science WW-4, LKO, University of Erlangen-Nuremberg, Martensstrasse 7, 91058 Erlangen, Germany;*

[2]*SSRL Materials Science Division, SLAC National Accelerator Laboratory, Menlo Park, California 94025, United States;*

[3]*Department of Materials Sciences 6, iMEET, University of Erlangen-Nuremberg, Martensstrasse 7, 91058 Erlangen, Germany;*

[4] *Institute of Biomaterials, University of Erlangen-Nuremberg, 91058 Erlangen, Germany;*

[5]*Department of Chemistry, King Abdulaziz University, Jeddah, Saudi Arabia*

*\*Corresponding author. Tel.: +49 91318517575, fax: +49 9131 852 7582*

Email: schmuki@ww.uni-erlangen.de

Link to the published article:

https://aip.scitation.org/doi/full/10.1063/1.4976010

In this study, we investigate noble metal free photocatalytic water splitting on natural anatase single crystal facets and on wafer slices of the [001] plane before and after these surfaces have been modified by high pressure hydrogenation (HPH) and hydrogen ion-implantation. We find that on the natural, intact low index planes photocatalytic $H_2$ evolution (in absence of noble metal co-catalyst) can only be achieved when the hydrogenation treatment is accompanied by the introduction of crystal damage, such as simple scratching, miscut in the wafer or by implantation damage. X-ray reflectivity (XRR), Raman, and optical reflection measurements show that plain hydrogenation leads to a ~ 1 nm thick black titania surface layer without activity, while a colorless, density modified and ~ 7 nm thick layer with broken crystal symmetry is present in the ion implanted surface. These results demonstrate that i) the H-treatment of an intact anatase surface needs to be combined with defect formation for catalytic activation, and ii) activation does not necessarily coincide with the presence of black color.

In photocatalytic $H_2$ generation, electron-hole pairs are created by illumination of a suitable semiconductor that is exposed to an aqueous solution; subsequent transfer of the excited electrons from the semiconductor conduction band to $H_2O$ then leads to the evolution of $H_2$. Due to its suitable energetic positions of band edges, high stability, and economic aspects, $TiO_2$ has been the most studied material over the last 40 years. Recently, so called 'black' $TiO_2$ has attracted wide attention for the photocatalytic generation of hydrogen [1-5]. Specifically, a large body of work addresses the high photocatalytic hydrogen evolution capability when 'black' titania is decorated with an adequate noble metal co-catalyst (Pt) [1, 5], which is needed to mediate electron transfer from the conduction band to the aqueous environment and as a catalyst for $H_2$ recombination.

Black $TiO_2$ was originally produced from anatase nanoparticles by a high pressure hydrogenation treatment at elevated temperatures. The black appearance was ascribed to a narrowing of the band-gap of anatase (3.2 eV) to a value of ≈1.5 eV, providing visible light absorption. A more recent very attractive finding is that titania crystallite powders, after high pressure hydrogenation, can form an intrinsic, stable co-catalytic feature for $H_2$ evolution [6, 7], i.e. no noble metal co-catalyst is needed to evolve $H_2$. Later reports also showed that titania powders or nanotubes, also after $H^+$ ion implantation [8] or an intense ball milling with $TiH_2$ [9], can show a similar intrinsic co-catalytic effect.

The majority of work (meanwhile several hundred reports on black $TiO_2$ have been published) uses polycrystalline anatase (powder, mesoporous structures, nanotubes) [1-9]. In contrast, we address in the present work the question whether specific crystal facets are of crucial importance for the creation of this intrinsic catalytic effect. It is worth noting that for anatase $TiO_2$, faceting has been reported to affect conventional photocatalytic reaction rates significantly; in general [001] planes are reported to be more reactive than [101] planes [10].

In our experiments, we used natural anatase crystals and polished [001] wafers obtained from SurfaceNet GmbH, Germany (as shown in Fig. 1). As natural anatase contains small amounts of impurities, giving the crystals a specific color (Mn typically red; Fe typically blue), we used both of these common crystals in our experiments (Fig. 1a). In a first set of experiments, the full crystals were hydrogenated at 500 °C at 20 bar, coated with epoxy resin except for the facet of interest, and immersed in a 50 v% MeOH solution ($CH_3OH$ serves here as hole capture agent) within a sealed quartz tube. Subsequently, the facet of interest was illuminated with a He-Cd laser (λ=325 nm, 50 mW, Kimmon, Japan) and hydrogen evolution was measured by gas chromatography and compared to non-hydrogenated crystals. To access the role of crystal defects, after the initial experiment, the corresponding crystal facet was intensely scratched using a diamond scribe, the crystal was hydrogenated again and the photocatalytic hydrogen measurement was repeated. The results of hydrogen evaluation for the different facets of anatase in the as grown ('intact') state and after scratching are compared in Fig. 1b. The results clearly show that for all intact facets no significant amount of hydrogen could be detected, whereas for all facets of the damaged crystals photocatalytic generation of H$_2$ could be observed.

In order to further elucidate the effect of defects, we miscut one of the anatase crystals approx. 10° with respect to the [101] plane (in the [001] direction) using a diamond saw. Also this miscut surface (or its corresponding defects) shows activity for photocatalytic H$_2$ evolution (Fig. 1b).

In order to clarify these findings and exclude effects (artifacts) of the naturally formed anatase facets, we additionally investigated wafer slides with an exposed [001] surface and used a HPH treatment. It is noteworthy that the high pressure hydrogenated wafer shows an optical color change to black (Fig. 1a). Nevertheless, also for these surfaces only traces of photocatalytic H$_2$ evolution could be measured. To introduce hydrogen and damage in a more defined way, we ion implanted the surface with H$^+$ at an energy of 30 keV with a dose of $10^{16}$ ions/cm$^2$ using a Varian 350 D ion implanter. These samples do not show any color change in the implanted region but a significant activation for photocatalytic H$_2$ evolution (Fig. 1b). Overall, these findings reveal that a high pressure hydrogenation treatment on (any of the investigated) 'intact' crystal surfaces does not yield the formation of an active co-catalytic center. Only hydrogenation in combination with damage (introduction of vacancy-interstitial pairs by ion implantation or the exposure of high indexed planes) can form such activating centers.

The most defined means of defect introduction, H-ion implantation, mainly leads to the formation of vacancy interstitial pairs with a depth-distribution that can be estimated using SRIM 2008 (Fig. 2a). According to these profiles, implant- and vacancy-contribution both peak at approx. 200 nm below the surface. In order to obtain experimental data on the alterations in the surface normal electron density caused by HPH and ion implantation, X-ray reflectivity (XRR) measurements were performed using 17.5 keV X-rays at a Bruker D8 reflectometer [15]. XRR measures $R(q_z)$, the interface-reflected intensity fraction of an x-ray beam incident on an interface at a grazing angle $\alpha$. $q_z = \left(\frac{4\pi}{\lambda}\right) \sin \alpha$ is the surface-normal wave vector transfer [16-18]. XRR is related, within the first Born-approximation, to the Fourier transform of the derivative of the surface-normal ($z$) electron density profile ($\rho(z)$) via the Master-formula [16-18]:

$$R(q_z) = R_\text{F}(q_z) \left| \rho_0^{-1} \int (\text{d}\rho(z)/\text{d}z) e^{-iq_z z} \text{d}z \right|^2 \quad .$$

$R_\text{F}(q_z)$ is the Fresnel reflectivity of an ideally smooth and abrupt interface. Detailed structural information can be obtained by constructing a model electron density profile, in the present case made up of slabs of constant electron density, which is then used to reproduce the experimental data by varying the parameters defining the model in the Master-formula.

Fig. 2b shows the Fresnel-normalized measured XRR (symbols) of an anatase single crystal before and after hydrogenation as well as after ion implantation, and the corresponding model fits (lines). The fit-derived electron density profiles are shown in Fig. 2c. The untreated sample shows the characteristic Fresnel fall-off of a single interface smeared by interfacial roughness. The model fit (red line) reproduces the data excellently, yielding a surface roughness of 4.8 Å. The Kiessig fringes observed for the HPH and H-implantation samples are typical for a thin surface layer with a different electron density profile than the underlying substrate. Upon hydrogenation, the model refinement and corresponding density profile (green line) in Fig. 2b and c indicate the formation of a 14 Å thick layer with a ≈10% reduced electron density. Ion implantation (blue line) on the other hand shows the presence of a 66 Å thick layer with an electron density of ≈80% of the bulk to be present at the surface of the implanted wafer.

Additionally, Raman spectra were acquired for the single crystal samples before and after HPH treatments and H-implantation (Fig. 3a). Particularly notable is a relative increase of the Eg bands at 144 and 636 cm$^{-1}$ relative to the B1g and the A1g peaks after implantation. This indicates a break of the symmetry of the (001) plane [14]. This is in contrast to HPH crystals that show all the typical intact anatase peaks in the Raman spectra with an intensive distribution in line with the non-hydrogenated layer. Obviously, the modified layer by H-implantation causes a much stronger co-catalytic effect than plain hydrogenation – however, as shown in Fig. 1a and 3b, implantation does not significantly alter the optical properties (the optical gap), which is in contrast to hydrogenation where typical spectra of black $TiO_2$ given in Fig. 3b show a strong tail into the visible range. These tails have been ascribed to various origins, such as hydrogenated amorphous layers or the formation of $Ti^{3+}$ species. From the present results one may, however, conclude that the formation of disorder and crystal damage by vacancies, interstitials, or the exposure of high index crystal planes is much more important than any alternation in the visible light absorption properties (i.e. blacking) for establishing photocatalytic $H_2$ evolution activity.

In conclusion, we show that on a defined intact (low index) single crystal surface of anatase, a $H_2$ treatment is not sufficient to create a stable activation for photocatalytic $H_2$ evolution. Clearly, additional lattice distortion (defects, exposed high index planes) are required to create an intrinsic co-catalytic effect.

ACKNOWLEDGMENT

We would like to acknowledge the ERC, the DFG, and the Erlangen DFG cluster of excellence (EAM) for financial support.

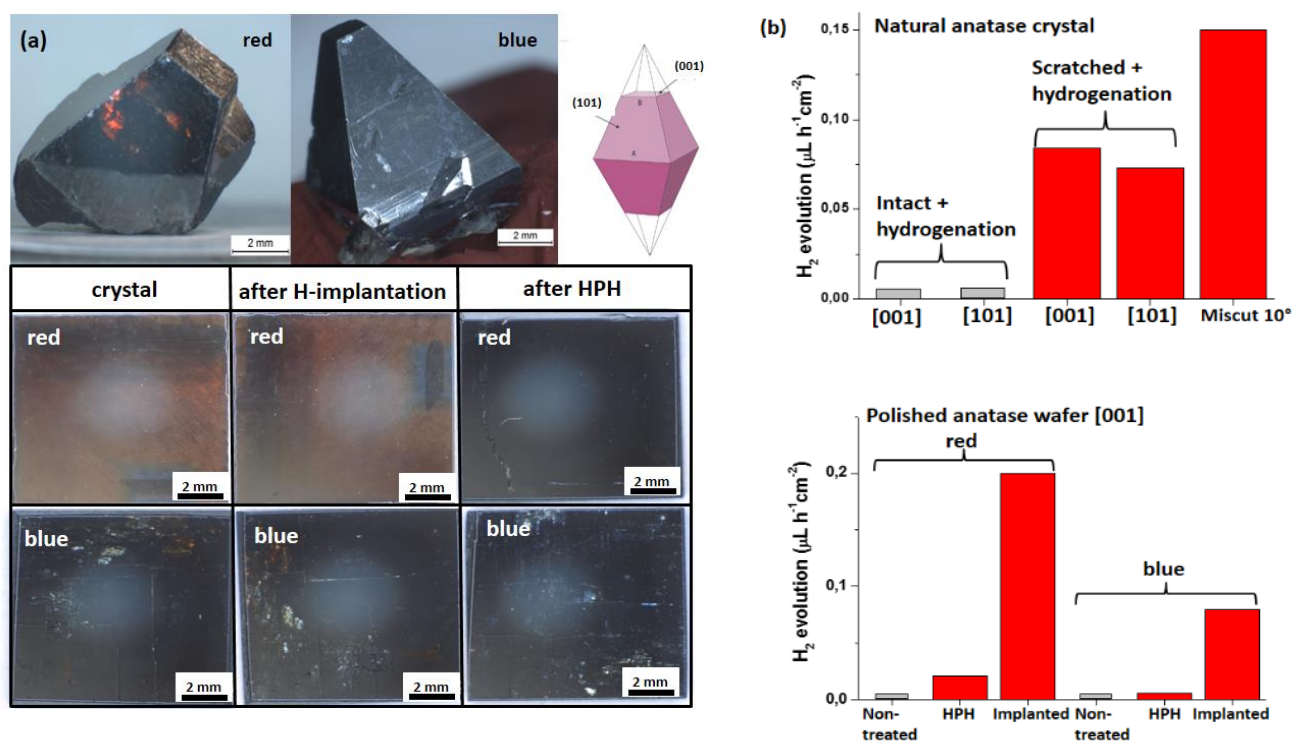

**Fig. 1** (a) Optical images of natural intact anatase single crystals and polished [001] wafers (red and blue, respectively) before and after HPH treatments and H-ion implantation; (b) photocatalytic hydrogen evolution rate under a He-Cd laser (λ=325 nm, 50 mW) illumination for natural anatase crystals (with different treatments: scratching, HPH and miscutting) and polished [001] wafer (before and after HPH and H-implantati

|  | red crystal | after HPH | after H-im |
|---|---|---|---|
| density substrate (e/Å³) | 1.15 | 1.15 | 1.15 |
| roughness substrate (Å) | 4.87 | 3.94 | 3.89 |
| density layer (e/Å³) | - | 0.86 | 1.04 |
| thickness layer (Å) | - | 14.39 | 65.44 |
| roughness layer (Å) | - | 7.57 | 2.67 |

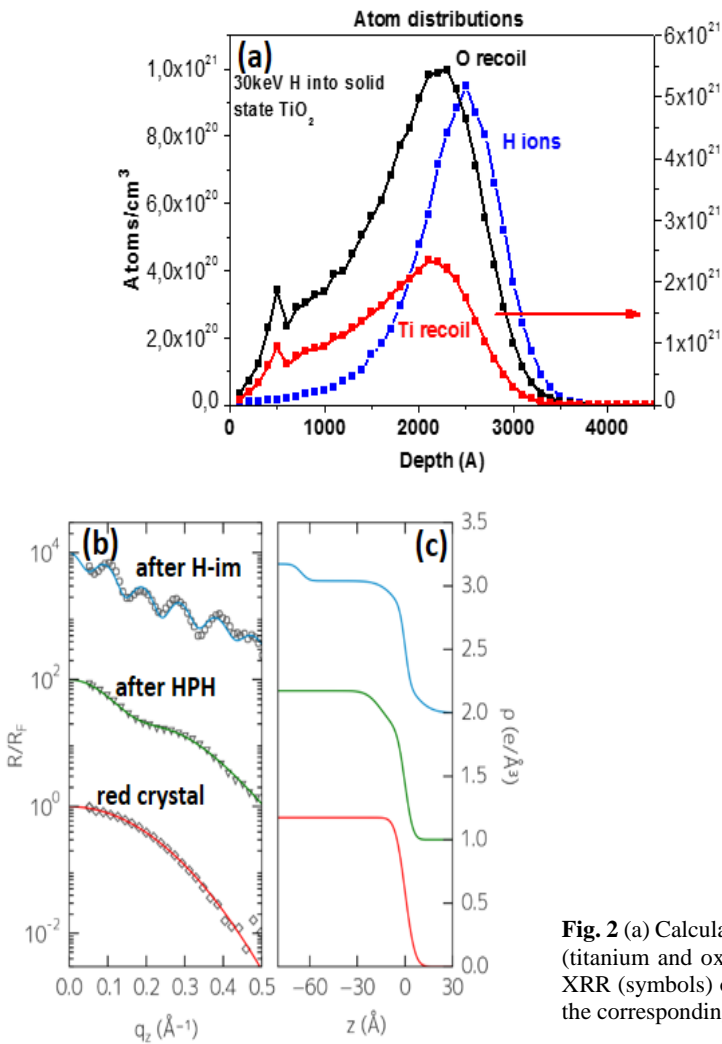

**Fig. 2** (a) Calculated depth distribution of implanted protons (H ions) and crystal damage (titanium and oxygen recoil) in a pure $TiO_2$ anatase substrate; (b) Fresnel-normalized XRR (symbols) of polished [001] wafers before and after HPH and H-implantation and the corresponding model fits; (c) fit-derived electron density profiles.

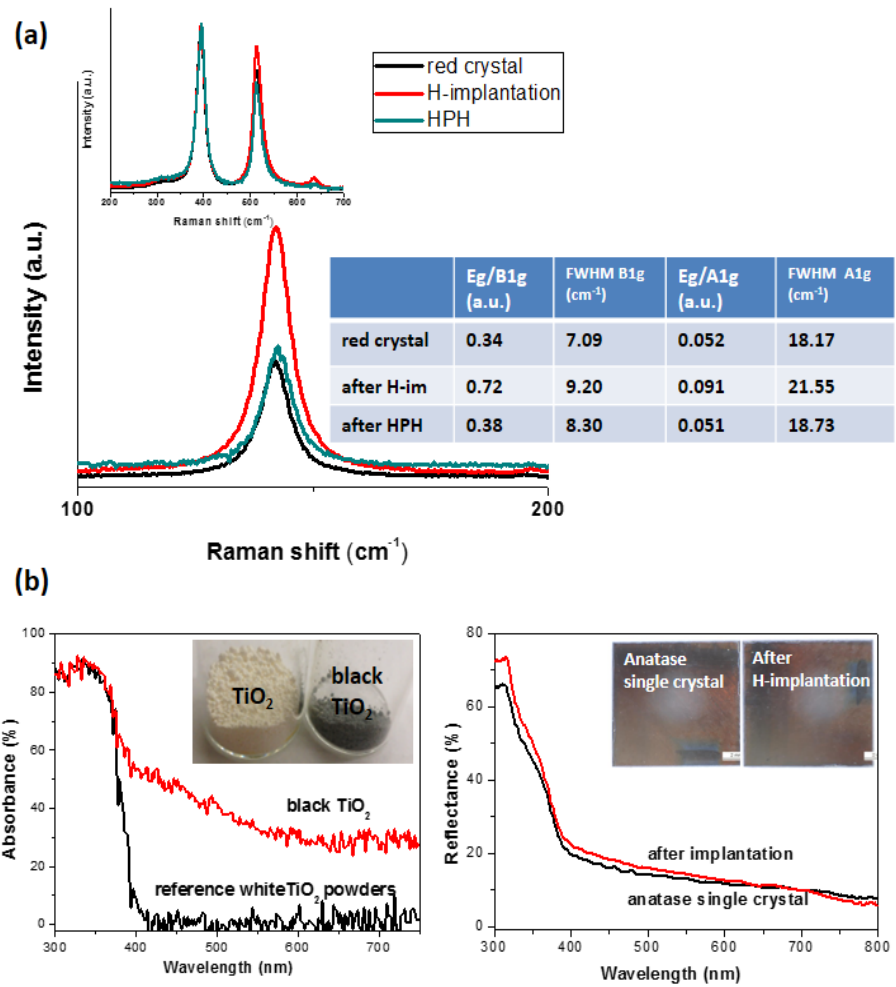

**Fig. 3** (a) Raman spectra of [001] anatase wafer before and after H-implantation and HPH; (b) integrated light reflectance spectra of black titania with reference white titania powders and polished [001] wafer before and after H-implantatio